\documentclass[12pt]{iopart}
\usepackage{graphicx,iopams}
\usepackage[normalem]{ulem}

\usepackage{color}

\begin{document}

\title[Individual bias and fluctuations in collective decision making]{Individual bias and fluctuations in collective decision making: from algorithms to Hamiltonians}

\author{Petro Sarkanych$^{1,2}$, Mariana Krasnytska$^{1,2,3}$, Luis G\'omez-Nava$^{4,5}$, Pawel Romanczuk$^{4,5}$, 
Yurij Holovatch$^{1,2,6,7}$}

\address{$^{1}$ Institute for Condensed Matter Physics of the National Academy of Sciences of Ukraine, 79011 Lviv, Ukraine}

\address{$^{2}$ $\mathbb{L}^4$ Collaboration \& Doctoral College for the
Statistical Physics of Complex Systems,
Leipzig-Lorraine-Lviv-Coventry, Europe}

\address{$^{3}$  Laboratoire de Physique et Chimie Th\'eoriques, Universit\'e de
Lorraine, BP 70239, 54506 Vand\oe uvre-les-Nancy Cedex, France}

\address{$^{4}$ Institute for Theoretical Biology, Department of Biology, Humboldt Universit\"at zu Berlin,
10099 Berlin, Germany}

\address{$^{5}$  Research Cluster of Excellence ``Science of Intelligence", 10587 Berlin, Germany}

\address{$^{6}$ Centre for Fluid and Complex Systems, Coventry University, Coventry,
CV1 5FB, United Kingdom}

\address{$^{7}$ Complexity Science Hub Vienna, 1080 Vienna, Austria}
 
\ead{petrosark@gmail.com}
\vspace{10pt}


\begin{abstract}
In this paper, we reconsider the spin model suggested recently to understand some features 
of collective decision making among higher organisms [A.T. Hartnett \emph{et al.}, Phys. Rev. Lett. {\bf 116} (2016) 038701]. 
Within the model, the state of an agent $i$ is described by the pair of variables corresponding to its opinion 
$S_i=\pm 1$ and a bias $\omega_i$ towards any of the opposing values of $S_i$.  Collective decision making 
is interpreted as an approach to the equilibrium state within the non-linear voter model subject to a social 
pressure and a probabilistic algorithm.  Here, we push such physical analogy further and give the statistical physics interpretation 
of the model, describing it in terms of the Hamiltonian of interaction  and looking for the equilibrium state via explicit calculation of its partition function. We show that depending on the assumptions about the nature of social interactions two different Hamiltonians can be formulated, which can be solved with different methods. In such an interpretation the temperature serves as a measure of fluctuations, not considered before in the original model. 
We find exact solutions for the thermodynamics of the model on the complete graph.  The general analytical predictions are confirmed using individual-based simulations. The simulations allow us also to study the impact of system size and initial conditions in the collective decision making in finite-sized systems, in particular with respect to convergence to metastable states.  
\end{abstract}

%
\vspace{2pc}

%
\submitto{\PB}
%
%
%
 
\section{Introduction} \label{I}

Collective decision making is omnipresent in biological systems. It can be observed across a wide range of scales ranging from cellular ensembles \cite{haeger2015collective}, via groups of social animals \cite{bazazi2011nutritional,strandburg2015shared} to entire societies or colonies \cite{ladyman2020complex,thurner2018introduction}. The ability of biological collectives to make accurate collective decisions, even when individuals have limited information about the state of the group and of the environment, inspired researchers across many disciplines, including physicist studying complex systems and self-organization \cite{holovatch2017complex, GomezNavaNatPhysFish}, or engineers interested in bio-inspired collective decision algorithms for artificial, distributed multi-agent systems {\cite{bizyaeva2022nonlinear,raoufi2021speed}}.

There have been significant advances in our understanding of collective decision making over the past decades, providing insights for example on the role of different interaction networks \cite{poel2021spatial}, correlated information\cite{winklmayr2023collective} or agent heterogeneity~\cite{GomezNavaNatPhysSheep}, many fundamental questions remain still open. On the theoretical side, idealized physics-inspired models provide a very valuable tool to investigate universal properties of collective decision making in very large systems and at large time-scales, where the microscopic details of the individual deliberation process and the interactions between agents can be ignored. On the one hand such models can be very efficiently numerically simulated, and on the other hand, even more importantly relying on  analogies to classical spin models in physics, they allow to employ analytical methods from statistical physics to deepen our understanding on the role of various factors. Recently, spin models have been used for example to model the decision making of animal groups on the move, and allowed to establish a bridge to neuronal decision-making within a single individual  \cite{pinkoviezky2018collective,sridhar2021geometry}. 

Recently, Hartnett \emph{et al.} \cite{hartnett2016heterogeneous} proposed a lattice spin model for a binary collective decision making task. Their main aim was  to investigate the role of heterogeneity in preferences as well as the role of unbiased individuals in collective decision making. The model and the corresponding study was motivated by previous empirical and theoretical work highlighting the unexpected impact of unbiased individuals for collective decision making in groups featuring individuals with conflicting biases to \cite{couzin334leonard}.

In this work, we want to push the physics analogy further: whereas Hartnett \emph{et al.} defined their model at the algorithmic level, by formulating an update rule for individual agents (spins) including coupling between agents via social field, we follow a more classical statistical physics approach to first formulate Hamiltonians for the system, which in turn enables us then to calculate partition functions and analyze the free energy landscape to identify steady-state solutions. 
A core difference of many collective decision making models to the classical spin models in physics is that the social coupling between agents (or spins) is not given by a (linear) superposition of pair-wise interactions, but typically is assumed to follow some non-linear response of the focal agent to an effective (local) social field, e.g. a threshold-like response. This in turn, in the derivation of macroscopic theories leads typically to emergence of infinite hierarchies of coupled multi-agent (multi-spin) terms, requiring some sort of closure. We show that this is also the case here, and that depending on the type of approximation made, we arrive at different Hamiltonians. 

In contrast to Hartnett \emph{et al.}, for the sake of analytical tractability we will focus here on the case of fully-connected graph. This allows the derivation of exact solutions for the free energy from a given Hamiltonian. However, we will compare selected analytical results with agent-based simulations, and our general approach also sets the stage for future investigation of different graph structures. The rest of the
paper is arranged as follows: in the next Section \ref{II} we describe the algorithm of Ref. \cite{hartnett2016heterogeneous}
and suggest possible  Hamiltonians of many-agent models that adhere certain features of this algorithm. In Section \ref{III} 
we derive exact expressions for the partition functions of the suggested models and discuss their equilibrium 
thermodynamic properties. These are compared and confirmed by our numerical simulations presented in Section \ref{IV}.
We end by conclusions and outlook in Section \ref{V}.

\section{Models: From the algorithm to Hamiltonians} \label{II}

In this section we briefly describe an algorithm of collective decision making in a group of biased individuals  (subsection \ref{II.1})
and suggest two spin Hamiltonians (subsections \ref{II.2}, \ref{II.3}, correspondingly) that
share certain features of this algorithm. Let us note from
the very beginning that the algorithm described in subsection \ref{II.1} defines the
local dynamical update that should lead the multi-agent system to the stable state. In the subsequent 
two sections we will be interested in the equilibrium properties of the stable state, leaving aside 
the way systems approaches the equilibrium. In turn, this enables one to introduce different static models
as we discuss in subsections \ref{II.2} and \ref{II.3}.

\subsection{The model and its algorithmic interpretation}\label{II.1}

The model suggested by Hartnett \emph{et al.}. \cite{hartnett2016heterogeneous} describes collective
decision making in an inhomogeneous population of $N$ individuals, that consist of 
three groups: two groups that favour conflicting opinions (the informed or biased 
individuals) and one group of uninformed individuals, that do not have any bias 
towards preferred outcome. The opinion of an individual is described
by a binary `spin' variable $S_i=\pm 1$, $i=1,...,N$. Each individual may 
or may not exhibit a bias regarding its preferred state. 
The bias of the $i$th individual is described by a variable $\omega_i$ that may attain three values 
$\{\omega_0, \omega_+, \omega_-\}$. These values correspond to unbiased
($\omega_i=\omega_0=1$), biased to +1 ($\omega_i=\omega_+$), and biased to -1
($\omega_i=\omega_-$) individual. It is assumed that individual biases
$\omega_0$, $\omega_+$, and $\omega_-$
are randomly and uniformly distributed with densities 
$\rho_0$, $(1-\rho_0)\rho_+$, and $(1-\rho_0)\rho_-$, correspondingly. 
 An approach to equilibrium is described
within a variant of a discrete time nonlinear voter model: it is considered that at each 
time step an individual is a subject of a local social field $h_i$ that originates from 
 its nearest neighbours and is distorted by individual's bias $\omega_i$:
\begin{equation}
\label{eq_social_field}
\label{2.1}
h_i=\frac{\omega_i n_i^+ - n_i^-}{\omega_i n_i^+ + n_i^-}\, ,    
\end{equation}
with $n_i^\pm$ being a number of the $i$th individual nearest neighbours with
opinion $+1$ or $-1$, correspondingly. In turn, the social field exerted on the individual
at a time instance $t$ probabilistically defines its state at time $t+1$: 
an individual in state $-1$ at time $t$ switches to the state $+1$ at time
$t+1$ with the probability $G_i$, whereas an individual in state $+1$ at time $t$ switches 
to the state $-1$ at time $t+1$ with the probability $1-G_i$. The probability function is chosen
to be:
\begin{equation}
\label{eq_switching_probability}
G_i=\frac{1}{2}\Big ( 1 + \frac{\tanh (b h_i)}{\tanh (b)}\Big )\, ,    
\end{equation}
and involves a non-linearity parameter $0\leq b \leq \infty$.\footnote{In the original formulation of Ref. \cite{hartnett2016heterogeneous}
this parameter is denoted as $\beta$. Here, we use a different notation to avoid misinterpretation with the temperature.}
For the limiting values of $b$, when the bias is absent (all $\omega_i=1$),  the probability function leads to
the classical voter model \cite{clifford1973model,redner2019reality} (at  $b=0$) or to the majority-rule model 
\cite{krapivsky2003dynamics} (at $b=\infty$), that describes, 
in particular a zero-temperature discrete time Ising model dynamics
\cite{glauber1963time} . Choosing intermediate values of $b$  allows one
to interpolate between these two familiar types of dynamics. Summarizing the above description, it is worth to mention,
that the model of Ref. \cite{hartnett2016heterogeneous} is implemented by the following algorithm:
\begin{itemize}
\item[1.] choose an initial configuration of variables $S_i$ and $\omega_i$ for all sites $i=1,\dots,N$;
    \item[2.] calculate a local social field $h_i$, Eq. (\ref{eq_social_field}) and probability function $G_i$ (\ref{eq_switching_probability}) 
    for all sites $i=1,\dots,N$ ;
    \item[3.] change states $S_i=-1$ to  $S_i=1$ with probability $G_i$, change states $S_i=1$ to  
    $S_i=-1$ with probability $1-G_i$;
    \item[4.] repeat steps 2 and 3 until an equilibrium state is reached.
\end{itemize}

So far, the model has been analysed by extensive
computer simulations on a square 2D lattice. Although the principal goal 
of these studies was to understand the collective behaviour that arises in 
animal groups and is influenced by many factors, the model description deliberately
concentrated on an impact of underlying global factors making it similar to those
used in statistical physics. To make this analogy even closer, below we will analyse
several Ising-like models that describe spin systems with inhomogeneities
that  mimic the above
described bias. As it will become evident, algorithmic
formulation of the original model may have different counterparts when formulated
in terms of many-particle Hamiltonians. Moreover, such an approach will allow us
to study influence of thermal fluctuations on collective behaviour in the spin systems 
under consideration, which may reveal an impact of noise on collective 
information processing in large groups. Although the models we will consider below
can be analyzed for any spatial arrangements of spins, we will concentrate on
the case of a complete graph, when each node is connected to all other nodes. Such 
a choice may correspond to the situations when agents are able to be in contact 
independently of their proximity in space, also it will enable us to get exact 
solutions for thermodynamics. 

\subsection{Biased Ising  model (bi-model)}\label{II.2}

An explicit assumption of the algorithm described in section \ref{II.1} is that 
the opinion states are shared with neighbours via interaction.
Let us proceed by looking on an equilibrium state of the system of
interacting agents, each being characterized by a pair of variables
$S_i,\omega_i$ that describe individual opinion (state) and bias towards 
this state.
As it was already
mentioned above, we will consider the case, when all individuals interact pair-vise, 
irrespective  on what is the distance between them. To this end, let us consider the 
following biased Ising model (bi-model) on a complete graph with the Hamiltonian:
\begin{equation}
\label{eq_hamiltonian}
H_{bi}=-\frac{2}{N}\sum_{i<j} \omega_i\omega_jS_iS_j\, .
\end{equation}
Here and below, the sums span over all $N$ nodes of the graph and the  Ising spins $S_i=\pm1$
correspond to the opinion states. Hamiltonian (\ref{eq_hamiltonian}) generalizes Ising model
on a complete graph (Kac model \cite{kac68mathematical,stanley1971phase})
incorporating dependence on random variables \cite{Krasnytska20,Krasnytska21}. 
Although within the static model Hamiltonian considered here one should not expect the one-to-one 
correspondence with the dynamic algorithm of subsection \ref{II.1}, we aim to give further
conceptual background to the notion of bias that plays central role in the algorithm.
In line with the Hartnett {algorithm}, we will assume  
$\omega_i$ to be a function of the opinion state, $\omega_i=\omega(S_i)$, in such a way that $S_i$ is preferred, provided this state coincides with the bias of the individual $i$. On contrary, when the value of the opinion state variable $S_i$ does not coincide with the individual's bias, the value of 
$\omega_i$  disfavours such state. This can be achieved assuming the following dependence:
\begin{equation}
\label{eq_omega_S_i}
 \omega (S_i) = 1 + k_iS_i\, ,    
\end{equation}
where $k_i$ attains one of three values:
\begin{equation}
\label{eq_k_i}
k_i = \left \{ \begin{array}{ll}
 \epsilon_+\, , & \quad \mbox{biased to +1},
\\
\epsilon_0\,  ,
&
\quad \mbox{unbiased},
\\
  \epsilon_-\, ,
&
\quad \mbox{biased to -1}\, ,
\end{array}
\right.
\end{equation}
and $\epsilon_+>0$,  $\epsilon_0=0$, and  $\epsilon_-<0$, are model 
parameters that govern the strength of a bias.
In the model of section \ref{II.1} it is assumed that individual biases
$\omega_0$, $\omega_+$, and $\omega_-$
are randomly and uniformly distributed with respective densities
$\rho_0$, $(1-\rho_0)\rho_+$, and $(1-\rho_0)\rho_-$. 
This corresponds to the case when 
$k_i$ are i.i.d. random variables with a distribution
function:
\begin{equation}\label{2.6}\label{eq_k_dist}
P(k)=\rho_0\delta(k-\epsilon_0) + (1-\rho_0)\rho_+\delta(k-\epsilon_+) +
(1-\rho_0)\rho_-\delta(k-\epsilon_-)\, .
\end{equation}
Furthermore, we will assume that $\{k_i\}$ are randomly distributed and
fixed in a certain configuration. This assumption is quite natural and mimics the
fact that individual bias does not depend on individual location and does not change
in time. Such situation corresponds to the so-called 'quenched disorder' \cite{Brout59}.
The bi-model will be further analyzed below, in section \ref{III.1}.

\subsection{Non-interacting spins in a social field (sf-model)}\label{II.3}
To account for a bias on an agent state we have introduced in the Hamiltonian (\ref{eq_hamiltonian}) \
a pair interaction  between biased individuals.
Another approach to  the model 
described in subsection \ref{II.1} is to consider a system
of non-interacting spins $S_i$ each being a subject of
an inhomogeneous local social (magnetic) field $h_i$ (sf-model) with the Hamiltonian: 
\begin{equation}\label{2.7}\label{eq_Hamiltonian_local}
H_{sf}=-\sum_{i=1}^{N} h_iS_i\, ,
\end{equation}
where the local field $h_i$ is given by Eq. (\ref{eq_social_field}). 
However, the caution expressed above about the correspondence between the model Hamiltonian 
and the algorithm of section \ref{II.1} is to place here too. Indeed,
the notion of the `social field' (\ref{eq_social_field})  is implemented
in the algorithm via the dynamic update rule (\ref{eq_switching_probability}). Therefore,
strictly speaking there is no one-to-one  correspondence between the 
`social fields' of both cases. To proceed further with an explicit expression for $h_i$,
since the model is considered on a complete graph, one makes use of the following relations
\begin{equation}\label{2.8}\label{eq_n_i}
    n_{i}^{\pm} =\sum_{j\neq i} \delta_{S_j,\pm1}=\sum_{j=1}^N \delta_{S_j,\pm1}-\delta_{S_i,\pm1}=N_{\pm}-\delta_{S_i,\pm1},
\end{equation}
where $N_{\pm}$ denote the number of spins up or down respectively and $\delta$ is the Kronecker symbol.
There is an obvious normalization condition $N_{+}+N_{-}=N$. 
The order parameter (mean magnetization per site) reads:
\begin{equation}\label{2.9}\label{eq_m_def}
    m = \frac{1}{N}(N_{+}-N_{-})=\frac{1}{N}\sum_{j=1}^N S_j \in [0,1].
\end{equation}
Hence
\begin{equation}\label{2.10}\label{eq_N_pm}
    N_{\pm}=N\frac{1\pm m}{2}.
\end{equation}
In terms of the order parameter $m$, one can rewrite the local magnetic field in a more compact way:
\begin{equation}
\label{2.11}
\label{eq_local_field_1}
h_i=\frac{m(\omega_i+1) +\omega_i-1-\frac{2}{N}(\omega_i \delta_{S_i,1}-\delta_{S_i,-1})}{m(\omega_i-1)+\omega_i+1-\frac{2}{N}(\omega_i \delta_{S_i,1}+\delta_{S_i,-1})}\, .
\end{equation}
In the thermodynamic limit $N\to\infty$, terms proportional to $\frac1N$ can be neglected leading to the local 
magnetic field
\begin{equation}\label{2.12}\label{eq_local_field_2}
h_i=\frac{m(\omega_i+1) +\omega_i-1}{m(\omega_i-1)+\omega_i+1}\, .
\end{equation}

Accordingly,  Eq. (\ref{2.7}) is the Hamiltonian of a system of non-interacting spins in random local magnetic
fields $h_i$ (\ref{2.12}). The fields are functions of random variables $\omega_i$ 
that attain values $\{\omega_+,\, \omega_0, \, \omega_-\}$
with given distribution function (\ref{2.6}). Depending on the bias, the fields  
can attain one of three values:
\begin{equation}\label{2.13}\label{eq_hp_hm}
h_+ = \frac{(\omega_+ + 1)m+\omega_+-1}{(\omega_+ - 1)m+\omega_++1} ,
\hspace{1em}  h_0=m, \hspace{1em} 
h_- = \frac{(\omega_- + 1)m+\omega_--1}{(\omega_- - 1)m+\omega_-+1} 
\, .
  \end{equation}
Thermodynamics of the sf-model
will be considered below in subsection \ref{III.2}.

\section{Equilibrium state and macroscopic observables}\label{III}

In this section we obtain exact solutions for equilibrium thermodynamic
properties of models with Hamiltonians suggested above in subsections 
\ref{II.2} and \ref{II.3}. 

\subsection{Exact solution for the bi-model }\label{III.1}
Substituting (\ref{eq_omega_S_i}) into (\ref{eq_hamiltonian}) we rewrite the Hamiltonian as:
\begin{eqnarray} \nonumber
H_{bi}=-\frac{2}{N}\sum_{i<j} \omega_i\omega_jS_iS_j =  -\frac{1}{N}\sum_{i\neq j} \omega_i\omega_jS_iS_j=\\
\label{3.1}
-\frac{1}{N}\sum_{i,j} \omega_i\omega_jS_iS_j + \frac{1}{N}\sum_i \omega_i^2\, .
\end{eqnarray}
Note that there is no restrictions on sums over $i,j$ in the first term of the 
last expression and we used that $S_i^2=1$  to derive (\ref{3.1}). This property of
the spin variable leads to further simplifications of the Hamiltonian. Indeed,
as long as $\omega_i$ linearly depends on $S_i$, initially the Hamiltonian (\ref{3.1}) contains three- and four-spin interactions: terms proportional to 
products of three and four $S_i$, correspondingly. However, taking into account the above mentioned property ($S_i^2=1$) one arrives at the following 
representation of the bi-model Hamiltonian (\ref{3.1}): 
\begin{equation} \label{3.2}
H_{bi}=-\frac{1}{N}\sum_{i,j}S_iS_j -2 \langle k \rangle \sum_i S_i
-N \langle k \rangle^2 + 1 + \frac{2}{N}\sum_{i} k_iS_i + 
\langle k^2 \rangle\, ,
\end{equation}
where $\langle k \rangle = \frac{1}{N}\sum_i k_i$   and $\langle k^2 \rangle = \frac{1}{N}\sum_i k_i^2$   are the mean and mean square of the random variable $k$.
Note that the Hamiltonian (\ref{3.2}) is that of the Ising model in a  local external field.

To analyse the thermodynamic properties, one defines the partition function for a given
configuration of random variables $\{k\}$:
\begin{equation} \label{3.3}
Z_{bi}(\{k\})= {\rm Sp}\, e^{-\beta H_{bi}}\, , \hspace{2em} {\rm Sp}(\dots)=\prod_i\sum_{S_i=\pm 1} (\dots)\, ,
\end{equation}
and $\beta=1/(k_BT)$.
In a standard setting, the next step is to define the configuration-dependent
free energy $G_{bi}(\{k\})=-\beta^{-1} \ln Z_{bi} (\{k\})$ and only then to perform an averaging
with distribution function (\ref{2.6}). However, as we will see below, considering model
on a complete graph essentially facilitates the problem. To proceed further, one makes use of the
Stratonovich-Hubbard  transformation writing for the first term in the Hamiltonian (\ref{3.2}):
\begin{equation} \label{3.4}
e^{\frac{\beta}{N}\sum_{i,j}S_iS_j}= e^{\frac{\beta}{N}\big (\sum_{i}S_i)^2\big ) } =
\sqrt{\frac{N}{4\pi\beta}} \int_{-\infty}^{+\infty}{\rm d}x\, e^{\frac{-N}{4\beta}x^2 + x \sum_iS_i}\, ,
\end{equation}
whereas for the whole Hamiltonian one gets:
\begin{equation} \label{3.5}
e^{-\beta H_{bi} } =
\sqrt{\frac{N}{4\pi\beta}} e^{\beta N \langle k \rangle ^2 - \beta \langle k^2 \rangle - \beta} \int_{-\infty}^{+\infty}{\rm d}x\, e^{\frac{-N}{4\beta}x^2}
\prod_i e^{f(x,k_i)S_i}\, ,
\end{equation}
with
\begin{equation} \label{3.6}
f(x,k_i) = x + 2 \beta \langle k \rangle - \frac{2\beta}{N}k_i\, .
\end{equation}
Now it is straightforward to take trace (\ref{3.3}) and to get for the partition function:
\begin{equation} \label{3.7}
Z_{bi}\simeq  \int_{-\infty}^{+\infty}{\rm d}x\, e^{\frac{-N}{4\beta}x^2} 
e^{\sum_i \ln\cosh f(x,k_i)}\, .
\end{equation}
Here and below we omit factors irrelevant for the subsequent analysis.
Substituting for  $N\to \infty$ the sum over all sites  by the sum over all
values of $k$:
$$
\sum_{i=1}^N \ln\cosh f(x, k_i) = N \sum_{\{k\}} P(k) \ln\cosh f(x, k)\, ,
$$
(in our case $k$ spans three values (\ref{eq_k_i}) and $P(k)$ is given by (\ref{2.6}))
we arrive at the following expression for the partition function:
\begin{equation} \label{3.8}
Z_{bi}\simeq  \int_{-\infty}^{+\infty}{\rm d}x\, e^{-N g(x) } \, ,
\end{equation}
where
\begin{eqnarray} \nonumber
g(x) &=& \frac{1}{4\beta}x^2 - \rho_0 \ln\cosh (x + 2 \beta \langle k \rangle) - (1-\rho_0)\rho_+ 
\ln\cosh (x + \\ \label{3.9} && 
2 \beta \langle k \rangle -  \frac{2\beta}{N}\epsilon_+) -  
 (1-\rho_0)\rho_- \ln\cosh (x + 2 \beta \langle k \rangle - \frac{2\beta}{N}\epsilon_-)\, ,
\end{eqnarray}
and
\begin{equation} \label{3.10}
\langle k \rangle = (1-\rho_0)(\epsilon_+\rho_+ +\epsilon_-\rho_-)\, .
\end{equation}
This expression gives an exact solution for the partition function. 
Note that although the partition function was calculated for the fixed (quenched) sequence
of random variables $\{k\}$ the resulting expression does not depend on a particular sequence,
but rather of their mean values. This is a result of self-averaging, typical for random spin 
models on a complete graph \cite{Krasnytska20,Krasnytska21}.  In the thermodynamic limit 
$N\to \infty$ keeping the leading terms and taking into account that $\rho_+ +\rho_-=1$ one gets 
for the function (\ref{3.9}) :
\begin{equation} \label{3.11}
     g(x) = \frac{x^2}{4\beta} - \ln\cosh (x + 2 \beta \langle k \rangle)\, .
\end{equation}
With function (\ref{3.11}),  the integral (\ref{3.8})  has the usual form 
of the partition function of the Ising model in an external field on the complete
graph. It is a textbook exercise to take the integral by the steepest descent method 
getting the following expression for the Gibbs free energy per spin:
\begin{equation}\label{3.12}
\beta g(x_0)= - \lim_{N\to \infty} \frac{\ln Z_{bi}}{ N} = \frac{(x_0-2 \beta \langle k \rangle)^2}{4\beta} - \ln\cosh (x_0)
\end{equation}
with $x_0=x_0(\beta,\langle k \rangle$) being the coordinate of $g(x)$ minimum:
\begin{equation}\label{3.13}
\frac{{\rm d}\, g(x)}{{\rm d}\, x} |_{x=x_0} =0,
\hspace{3em} \frac{{\rm d}^2\, g(x)}{{\rm d}\, x^2} |_{x=x_0} > 0\, .
  \end{equation}
  
\begin{figure}[t!]
    \centering
    \includegraphics[width=0.3\textwidth]{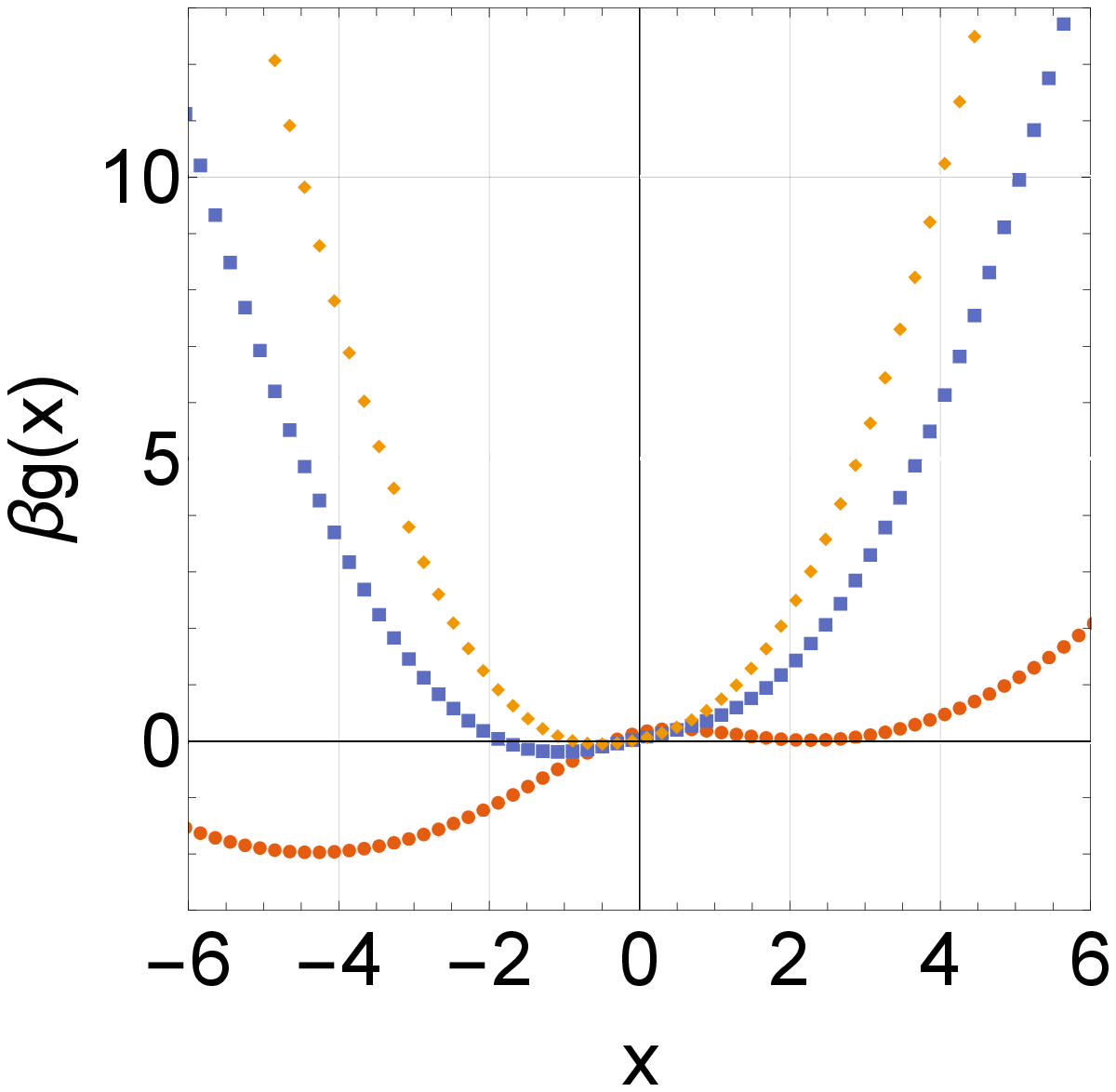}\hspace{0.1em}
    \includegraphics[width=0.3\textwidth]{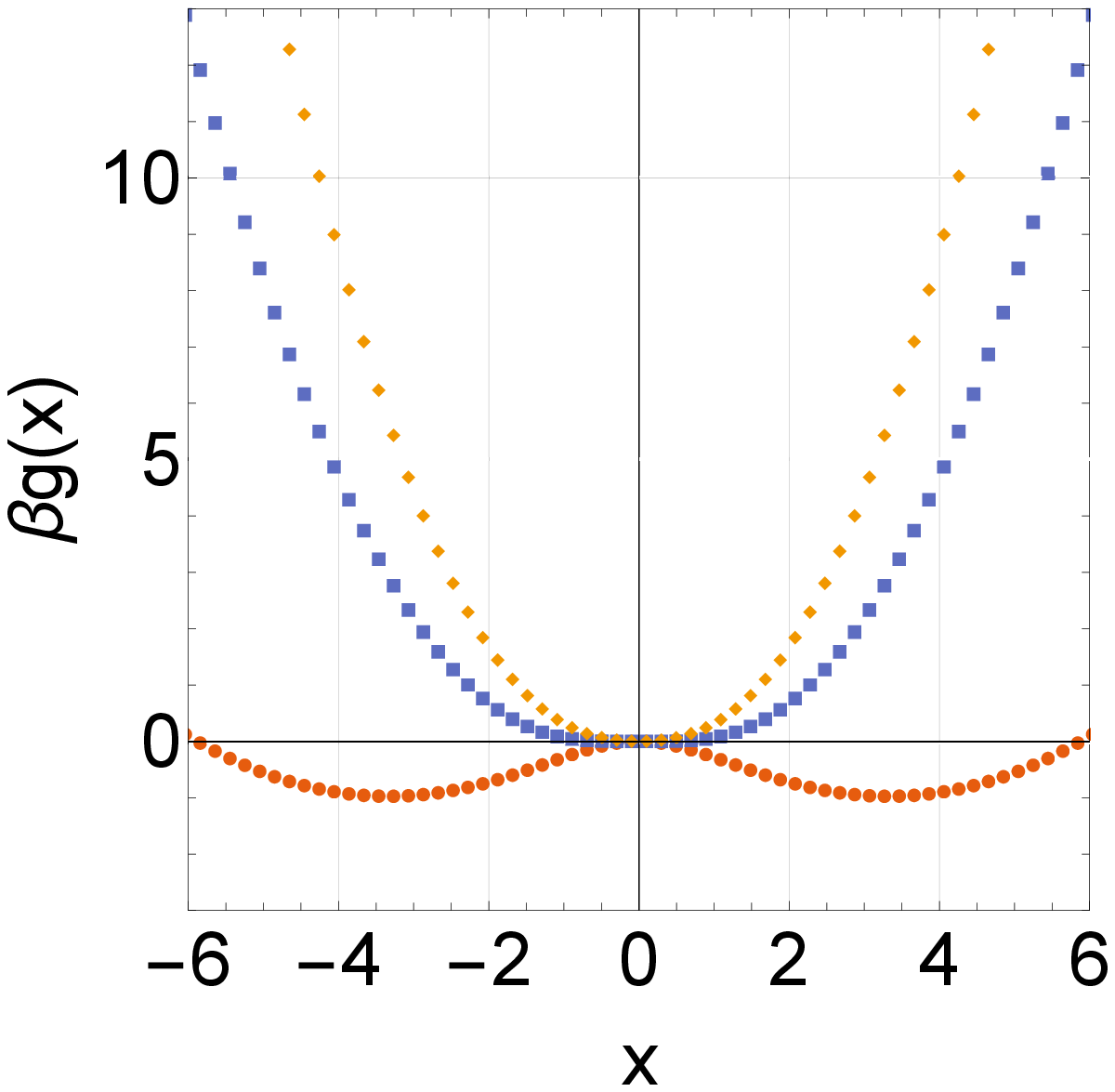}\hspace{0.1em}
    \includegraphics[width=0.36\textwidth]{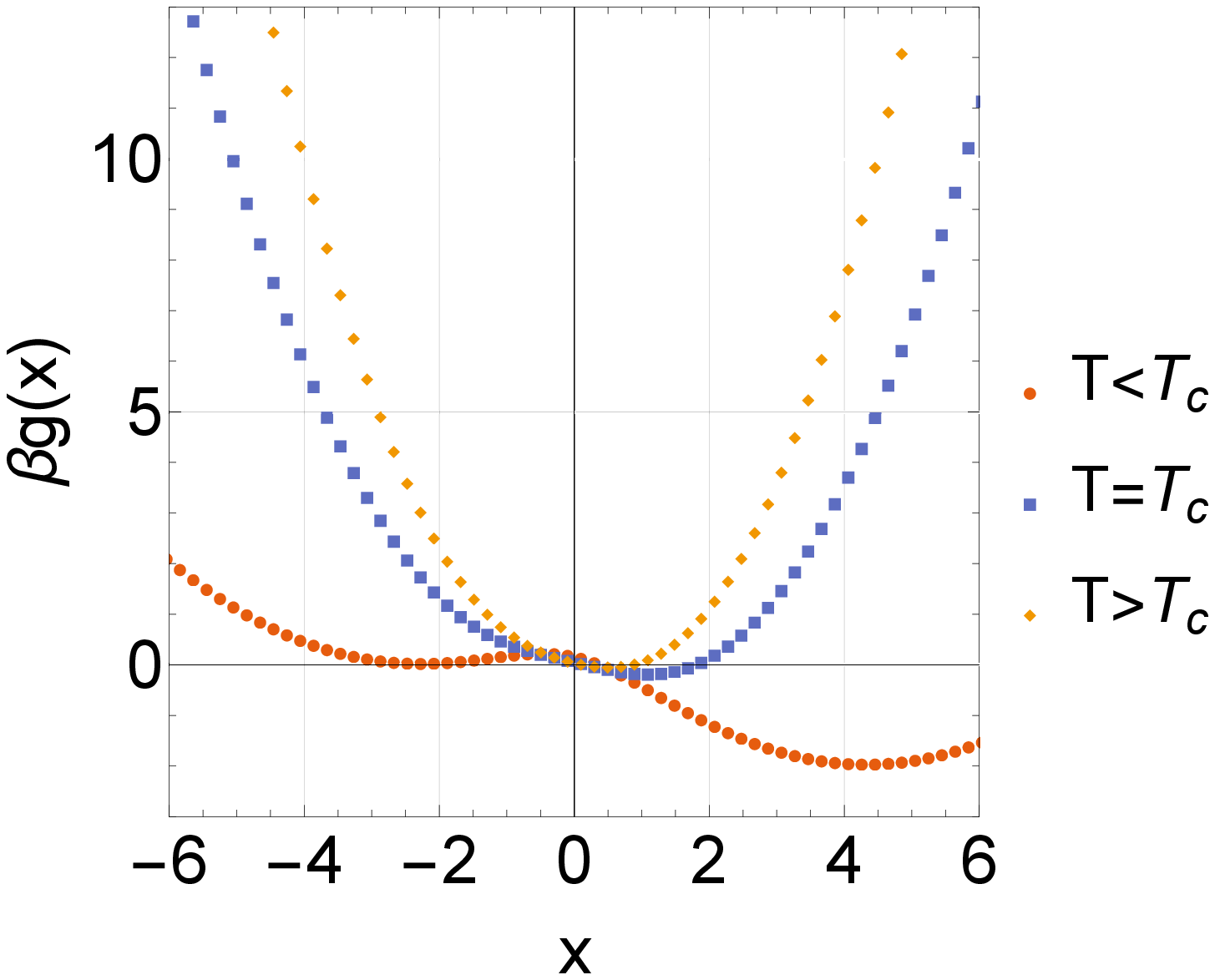} 
    \\
    { \hspace{-0.35cm}(a) \hspace{4.2cm} (b)} \hspace{4.2cm} (c)\\
    \caption{Function (\ref{3.12}) at fixed values   $\langle k \rangle=-0.3$ (a); $\langle k \rangle=0$ (b);
    $\langle k \rangle=0.3$ (c) and different temperatures  $T>T_c$, $T=T_c$, and $T<T_c$.} 
    \label{fig3.1}
\end{figure}

Typical behaviour of function (\ref{3.12}) is shown in Fig. \ref{fig3.1}. 
There, we plot $\beta g(x)$ for  $\langle k \rangle=-0.3$, $\langle k \rangle=0$,  
and $\langle k \rangle=0.3$ at different values of $T$. 
Obviously the second case corresponds to the absence of an external field (no bias), whereas the first and
the third one are symmetric counterparts. The critical temperature, separating two
regimes in Fig. \ref{fig3.1}b readily follows from  (\ref{3.11}) 
at $\langle k \rangle=0$: $\beta_c^{-1}=k_BT_c=2$. The first obvious observation in terms of the problem
considered here is that any non-zero value of $\langle k \rangle$  (any bias) leads to non-vanishing
value of $x_0$ at any finite temperature $T$. As it follows from Eq. (\ref{3.10}), the  value 
$\langle k \rangle=0$ is achieved either when all individuals are unbiased ($\rho_0=1$) or for equal mean strengths
of oppositely biased individuals ($\epsilon_+\rho_+ = - \epsilon_-\rho_-$). Another observation is that 
two minima are present at low temperatures $0\leq T \leq T_1$ (dotted red curves in Figs. \ref{fig3.1}a,c). 
Since the integral in (\ref{3.8}) is evaluated by the steepest descent method,
only the global minimum contributes to the free energy (\ref{3.12}) in the thermodynamic limit $N\to \infty$.
For the finite system size however, the local minimum contributes too and corresponds to the metastable state. 
In particular, such metastable states influence crossover to the stable state, see \cite{Bovier22} and references
therein for further discussions.
For the 
temperature $T_1$ at which the local minimum disappears one gets:
\begin{equation}\label{3.14}
T_1=T_c\Big(1-\Big[ \frac{9}{4} \langle k \rangle^2 \Big]^{1/3}\Big)\, .    
\end{equation}
Relation between $x_0$ and magnetization $m$ is given by the equation of state: 
\begin{equation}\label{3.15}
m(h,\beta)=- \Big( \frac{\partial g(\beta,h)}{\partial h} \Big )_\beta\, ,    
\end{equation}
where the Gibbs free energy density at the presence of an external magnetic field $h$ readily
follows from Eq. (\ref{3.12})
\begin{equation}\label{3.16}
g(\beta,h)=  \frac{[x_0(\beta,h)- \beta(2 \langle k \rangle +h)]^2}{4\beta^2} - \frac{\ln\cosh (x_0(\beta,h))}{\beta}\, ,
\end{equation}
with $x_0(\beta,h)$ being the solution of 
\begin{equation}\label{3.17}
\frac{x}{2\beta} -   \langle k \rangle - \frac{h}{2} -  \tanh x =0 \, .
  \end{equation}
Substituting (\ref{3.16}) into (\ref{3.15}) one arrives at the following relation between $m$ and
$x_0$:
\begin{equation}\label{3.18}
m(h,\beta) = \frac{x_0(\beta,h)}{2\beta}- \langle k \rangle - \frac{h}{2}\, ,
  \end{equation}
and the following expression for the free energy:  
\begin{equation}\label{3.18b}
g(\beta,h)=  m^2 - \frac{1}{\beta} \ln\cosh (2\beta m + \langle k \rangle + \frac{h}{2})\, .
\end{equation} 
The magnetization $m\equiv m(h,\beta)$ is found from the equation for the extremum (\ref{3.13}):
\begin{equation}\label{3.20}\label{eq_magnetization_TD}
m = \tanh \Big \{2 \beta (m + \langle k \rangle + \frac{h}{2})\Big \} \, .
  \end{equation}
 Note that the zero temperature solution of Eq. (\ref{3.17}) reads
\begin{equation}\label{3.19}
\lim_{\beta\to\infty}\frac{x_0(\beta,h)}{2\beta} =   1 +  \langle k \rangle + \frac{h}{2}\, ,
  \end{equation}
and leads to a proper normalization of the magnetization given by Eq. (\ref{3.18}): $m(\beta\to\infty,h)=1$.

\begin{figure}[t!]
    \centering
    \includegraphics[width=0.7\textwidth]{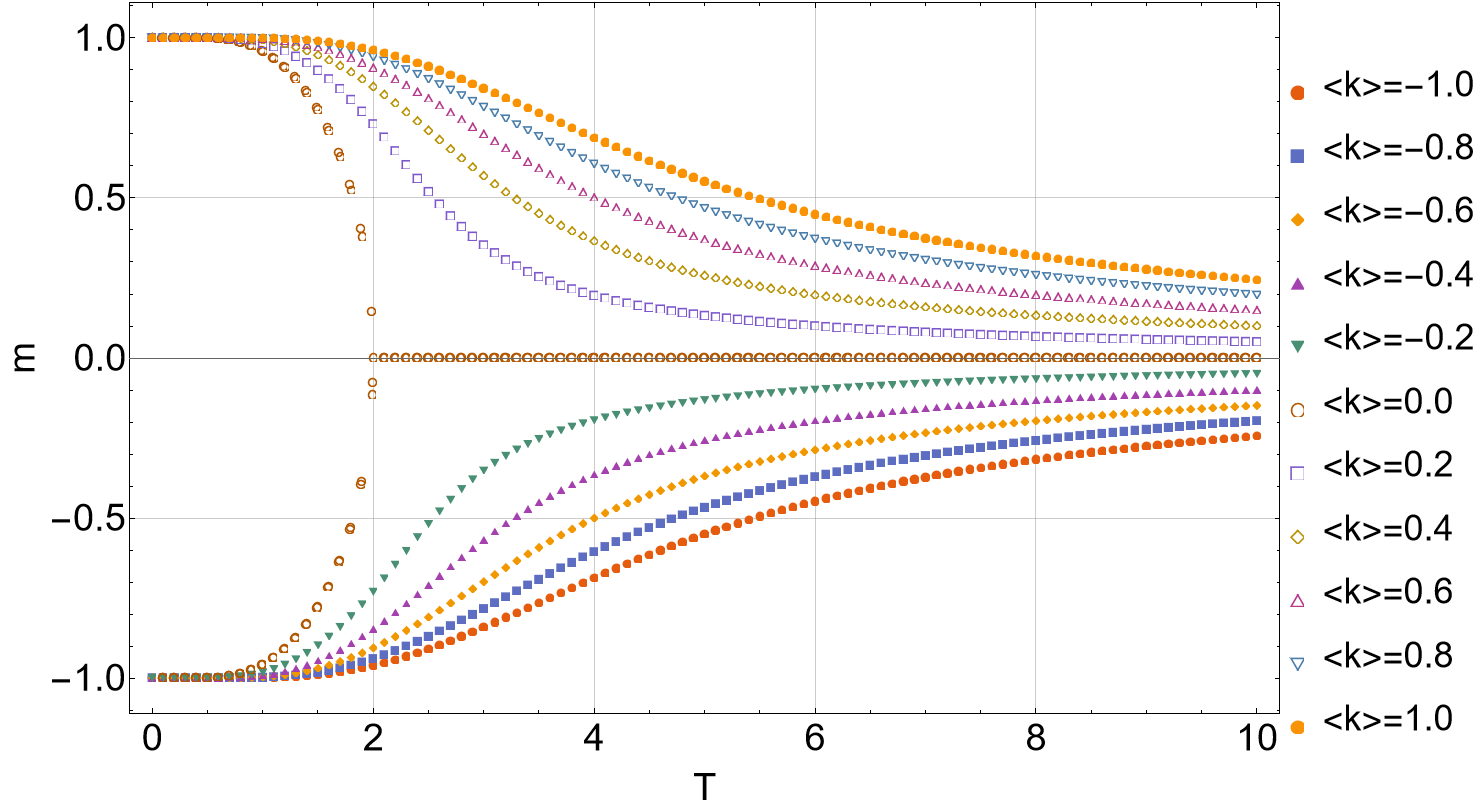}
    \caption{Temperature behaviour of the spontaneous magnetization $m(T,0)$ (\ref{3.20}) at 
    $-1 \leq \langle k \rangle \leq 1$.
    \label{fig3.2}}
    \label{fig_m_vs_T_TD}
\end{figure}

In Fig. \ref{fig3.2} we show the spontaneous magnetization $m(\beta,0)$ (\ref{3.20}) for different values of
$\langle k \rangle$ ranging from -1 to 1 with a step 0.2.
We will compare other features of 
considered here bi-model with those of the algorithmic model in section \ref{IV}.
 
\subsection{Exact solution for the sf-model}\label{III.2}

The  sf-model  partition function $Z_{sf}(\{\omega\})$ is related 
to the Hamiltonian $H_{sf}$  (\ref{2.7})
of the non-interacting
spins in random social field by taking trace over spins as in Eq. (\ref{3.3}). 
Similar as  it was shown in the former subsection for
$Z_{bi}(\{k\})$, the partition function $Z_{sf}(\{\omega\})$
is self-averaging with respect to the random variables $\omega$
leading to
\begin{equation} \label{3.21} \hspace{-5em}
Z_{sf}=2^N\exp\Big\{\sum_{i=1}^N \ln \cosh (\beta (h_i+h))\Big\}=
2^N\exp\Big\{\sum_{\omega} P(\omega) \ln \cosh (\beta (h(\omega)+h)\Big\}
\, ,
\end{equation}
where similar as in section \ref{III.1} we accounted for the homogeneous
external field $h$. Here, the  distribution function $P(\omega)$ is given by Eq. 
(\ref{2.6}), the summation in the last expression spans three values 
$\omega=\{\omega_0,\omega_+,\omega_-\}$, and the corresponding random 
fields $h(\omega)$ are given by Eq. (\ref{2.13}). In turn, the Gibbs
free energy per site reads:
\begin{eqnarray} \nonumber
- \beta g(\beta,h)&=& \rho_0\cosh(\beta (m+h)) + (1-\rho_0)\rho_+\cosh(\beta (h_++h)) + \\
&&
(1-\rho_0)\rho_-\cosh(\beta (h_-+h)) \label{3.22}
\, ,
\end{eqnarray}
and the magnetization $m(\beta,h)$ is found from the self-consistency relation. The last
relates mean magnetization that appears in the Hamiltonian (\ref{2.7}) with the mean spin value 
via:
\begin{equation}\label{3.22b}
m = \frac{1}{N}\Big \langle \sum_{i=1}^N S_i \Big \rangle \,
\end{equation}
where the averaging is performed over the Gibbs distribution with the Hamiltonian
(\ref{2.7}). Substituting (\ref{2.7}) into (\ref{3.22b}) one arrives at: 
\begin{eqnarray}\nonumber 
m(\beta,h) &=& \rho_0\tanh(\beta (m+h)) +  (1-\rho_0)\rho_+\tanh(\beta (h_+ + h)) + \\ &&
\label{3.23}
(1-\rho_0)\rho_-\tanh(\beta (h_-+h))
\, .
  \end{eqnarray}

\begin{figure}[t!]
    \centering
    \includegraphics[width=0.7\textwidth]{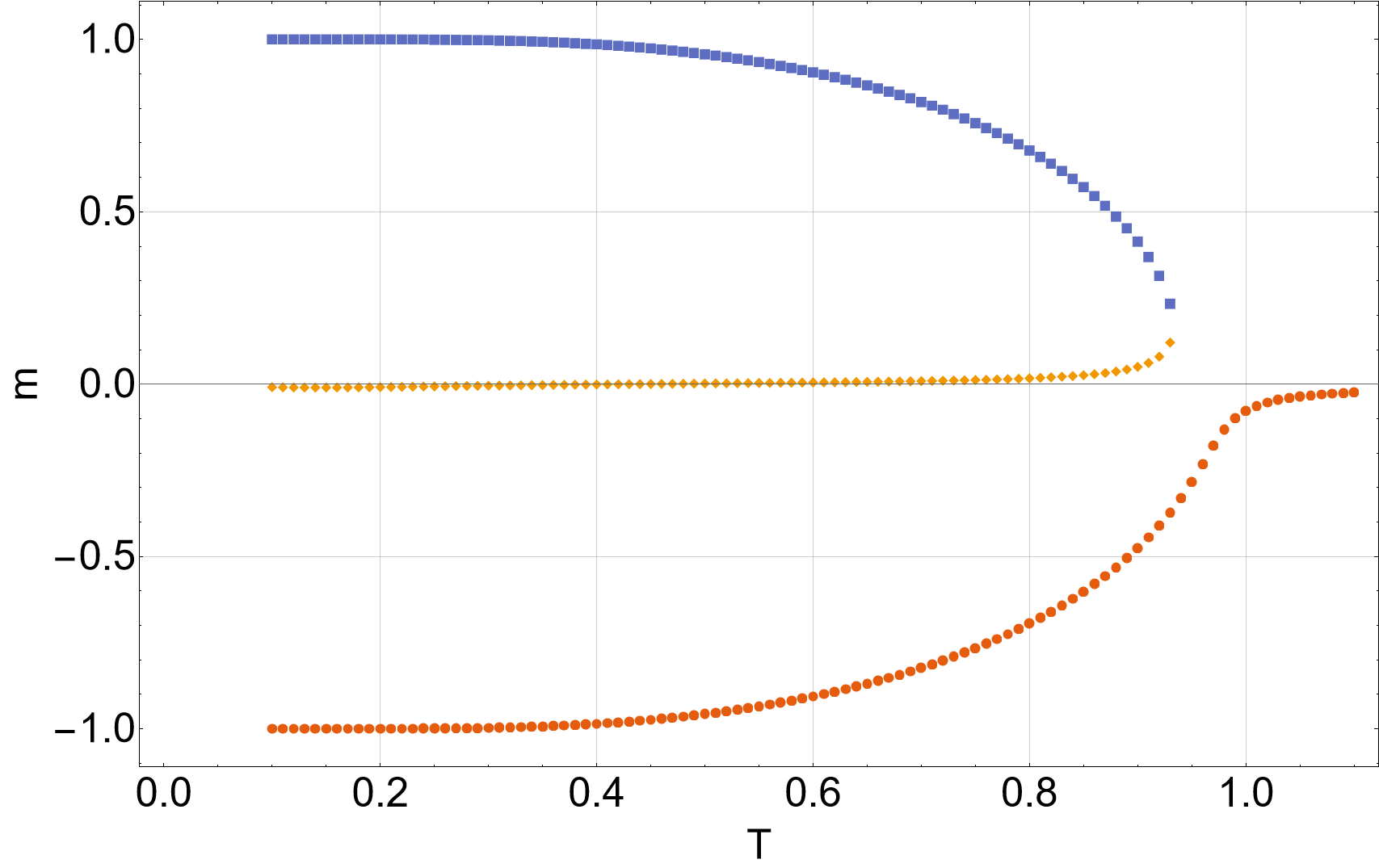}
    
    \caption{Solutions of the equation (\ref{3.23}) at fixed values of model parameters
    $\rho_0=0.7, \rho_+=0.6, \rho_-=0.4, \omega_+=1.5, \omega_-=0.5$. 
    Different colours represent three branches. The lowermost red branch corresponds to the stable state
    with minimal free energy.
         } 
    \label{fig3.3}
    \label{fig_self_consistency_result}
\end{figure}

 For the zero-temperature magnetization with no external field $h=0$, 
all functions $\tanh (x)$ in (\ref{3.23}) can be replaced by $\mathrm{sign} (m)$ leading to the equation 
$m=\mathrm{sign} (m)$ that has two solutions $m(\beta\to\infty)=\pm 1$.
Typical behaviour of the solutions of the Eq. (\ref{3.23}) as functions of temperature are shown 
in Fig. \ref{fig3.3} for $\rho_0=0.7, \rho_+=0.6, \rho_-=0.4, \omega_+=1.5, \omega_-=0.5$.
Depending on $T$ there might be up to 3 solutions. Each of them is shown with different colour in the
figure. The stable state corresponds to the solution giving the minimal value of the free energy. 
For the chosen values of model parameters, it appears to be described by the lowermost curve (red online) 
in Fig. \ref{fig3.3}. The value of the free energy for the uppermost blue curve is only slightly higher 
than for the red one. Therefore, it is reasonable to assume that the blue curve describes the 
metastable state.  For the case illustrated in Fig. \ref{fig3.3} the free energy for the metastable state 
is only about 1\% higher than in the stable state. This increases the probability that the system 
remains in the metastable state.
With an increase of temperature the values of all three solutions get closer with the limiting value
$m(T\to\infty )=0$. 


\begin{figure}[t!]
    \centering
    \includegraphics[width=0.7\textwidth]{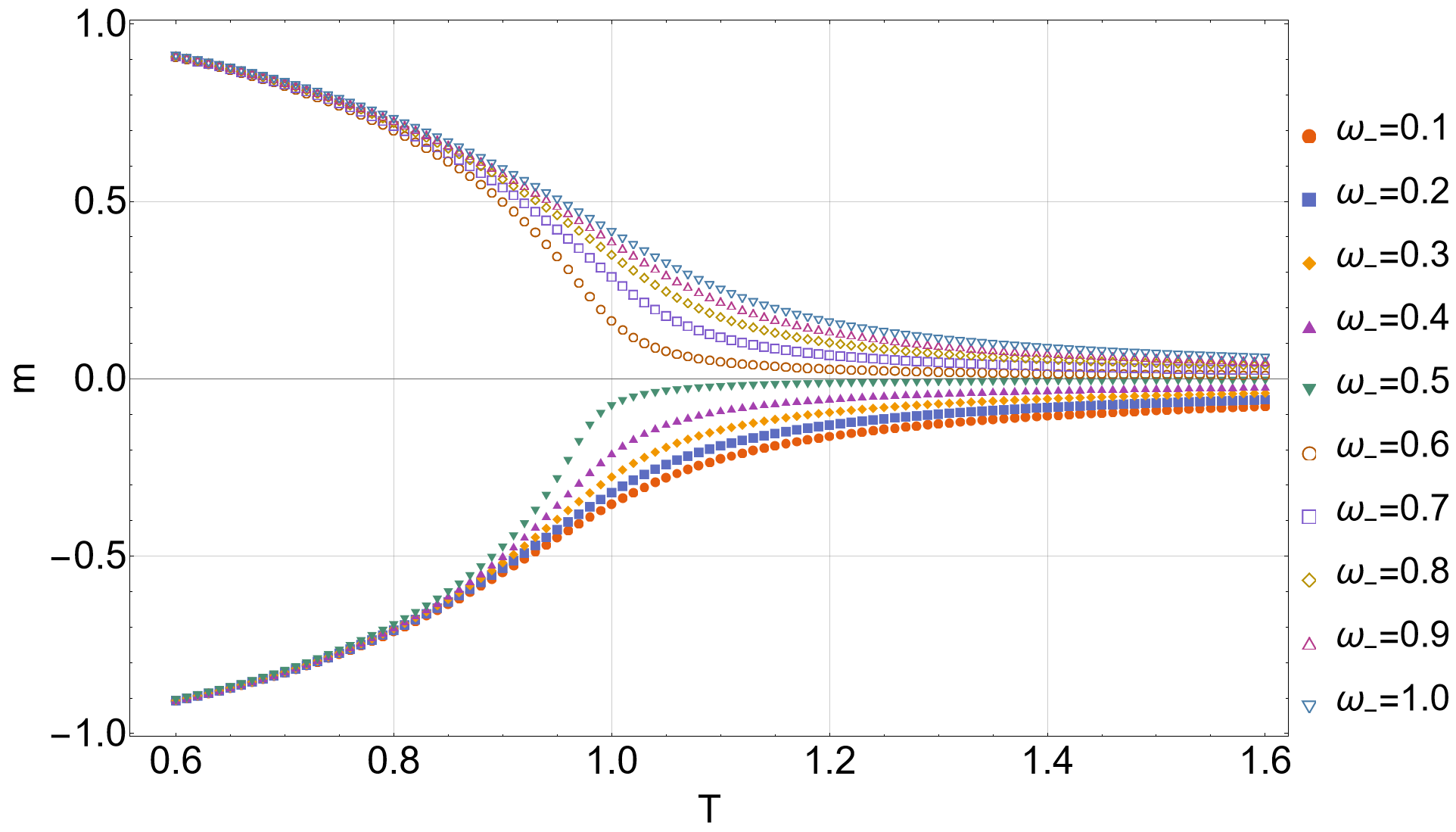}
    \caption{Stable state magnetization as a function of temperature at different values of $\omega_-$. The rest
    of the model parameters are the same as in Fig. \ref{fig3.3}.
     }
    \label{fig3.4}
\end{figure}

In Fig. \ref{fig3.4} we show the  stable state magnetization as a function of temperature 
 keeping the same set of parameters as in Fig. \ref{fig3.3} and choosing different bias 
 strengths $\omega_-$. Similarly to the bi-model we discussed in the previous 
 subsection, in the sf-model the magnetization remains non-zero at any finite temperature 
 (no transition is observed) and its sign and value depend on the parameters $(\rho_0, \rho_+, \rho_-, \omega_+, \omega_-)$. 

\section{Numerical simulations}\label{IV} 

In order to test the theoretical predictions, we performed numerical simulations of the individual-based model on the complete graph. To account for the finite temperature assumed in the theoretical analysis, we extend the Hartnett model by an additional random process that may induce state changes of individual agents: At each time step, irrespective of the social field exhibited, an agent will switch its current state from $\pm 1$ to $\mp 1$ with the probability $p_{noise}$. For $p_{noise}=0$, we recover the original Hartnett model on a complete graph. 

Please note, that there exist different possibilities to introduce noise into the system, and our choice was guided by simplicity and numerical convenience. However, there is no direct correspondence between the ``microscopic'' noise parameter $p_{noise}$ and the thermodynamic quantity $T$. Finally, for the extreme choice of $p_{noise}=1$, we will observe permanent switching of all agent states at each time step. Depending on the initial condition, e.g. an initial high consensus state this may result in a spuriously synchronized collective switching of the entire system, while maintaining consensus. Therefore, we restrict our analysis $p_{noise}\leq0.8$, which ensures randomization and vanishing consensus irrespective of initial conditions.

Whereas the dynamical evolution of the system certainly depends on the non-linearity parameter {$b$} in the Hartnett model,
see Eq. (\ref{eq_switching_probability}), the actual stationary states can be assumed in first approximation  to be independent of {$b$}. Therefore, we fix in our numerical simulation the nonlinearity parameter to  {$b=1$}.

\begin{figure}[t!]
    \centering
    \includegraphics[width=0.9\textwidth]{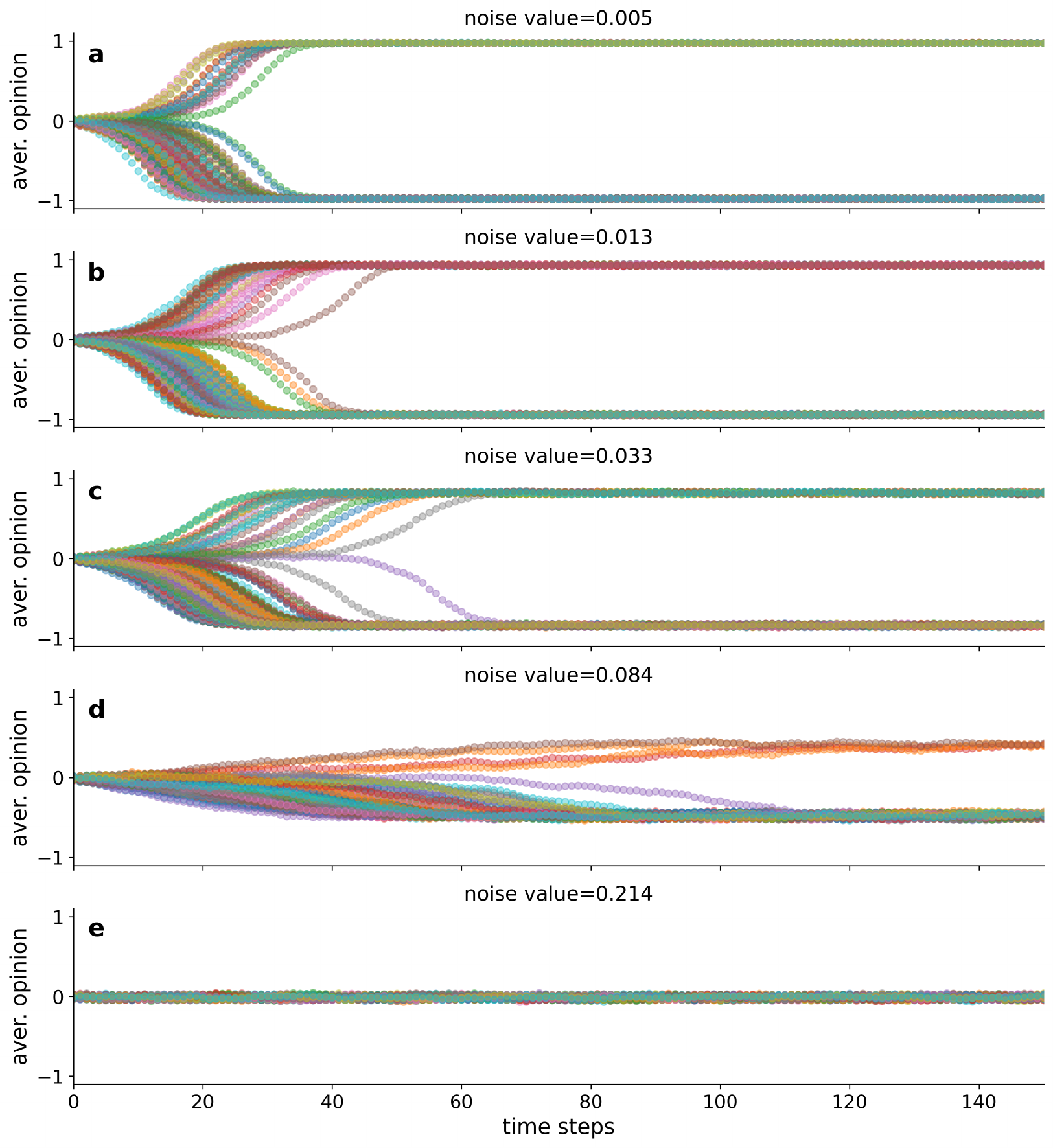}\hspace{3em}
    \caption{Average opinion of different simulations of a system of size $N=10 000$ over time. For these plots, we used the same parameters as Figure~\ref{fig3.3} for five different noise values. In each plot, we show 40 independent realizations where the system always satisfies the same condition: average opinion is zero. For all plots of this figure, we used initial conditions where half of the nodes of the network were in state $+1$ and the rest in state $-1$.}
    \label{fig5}
    \label{fig_avgm_vs_time}
\end{figure}

In Fig. \ref{fig5}, we show typical time courses of the average opinion (magnetization) of single simulations of the Hartnett model with noise on a complete graph. In general, for small $p_{noise}$ individual runs converge to a state of high consensus $|m|=1$ but not necessarily to the same average opinion $m$. Thus, averaged over many simulation runs we observe a bimodal distribution of final collective opinion states. For an unbiased system, e.g. with  $\rho_+=\rho_-=0$ or $\rho_+=\rho_-$, and {$\omega_+\omega_-=1$}, if we start from a fully disordered initial condition $m=0$, the average probability to observe the final state $\pm 1$ is 0.5. 

In a biased system, the results of the probability distribution of steady state at small $p_{noise}$ becomes asymmetric, with the opinion favored by the bias being more likely to be observed. However, finite-size fluctuations may always make individual simulation runs to converge to the steady state counter to the bias, in particular if the initial condition of the system is in a perfectly disordered state ($|m|(t=0)=0$).

With increasing noise $p_{noise}$ the consensus decreases, and we observe approach  towards $|m|=0$ in the limit of large $p_{noise}$ (see Fig. \ref{fig5} ). Due to finite fluctuations at large noise it is difficult to distinguish in numerical simulations a continuous phase transition predicted from the theory for the unbiased case from a finite, yet arbitrary small magnetization predicted in the biased case. In general, we observe a transition-like behavior with finite magnetization for small $p_{noise}$ and effectively vanishing magnetization at large $p_{noise}$.

\begin{figure}[t!]
    \centering
    \includegraphics[width=1\textwidth]{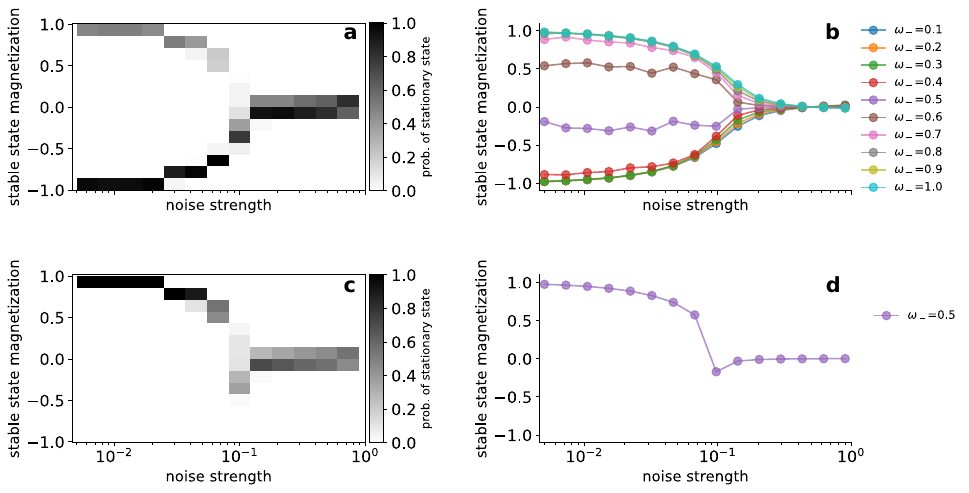}\hspace{3em}
    \caption{
    \textbf{a}. Results of the numerical simulations using the same parameters as Figure~\ref{fig3.3}. The results are presented as a two dimensional histogram (or heat-map) where, for each value of the noise strength, we computed 200 realizations and then computed the probability to observe a given value of the steady-state magnetization. The value of this probability is given in a black and white scale. The initial conditions {were} such that half of the nodes  were in state $+1$ and the rest in state $-1$.
    \textbf{b}. Average values of the steady-state magnetization for different values of the parameter $\omega_-$ computed over 200 realizations. The rest of the parameters are the same as Figure~\ref{fig3.4}. Note that the heat-map on the left is complementary to the purple curve [$\omega_- = 0.5$]. Thus, although we observe that the mean magnetization for $\omega_- = 0.5$ is close to zero, these average values arise from [almost] symmetric values observed in the heat-map. 
    The initial conditions {were} such that half of the nodes were in state $+1$ and the rest in state $-1$.
    \textbf{c}. Numerical results using the same parameters as subplot \textbf{a} but an initial condition where all the nodes in the network are in state $+1$.
    \textbf{d}. Numerical results for parameter $\omega_-=0.5$ using the same parameters as subplot \textbf{b} but an initial condition where all the nodes in the network are in state $+1$.
    For all plots in this figure we used a system of size $N = 1000$. }
    \label{fig6}
    \label{fig_stat_m_diff_w}
\end{figure}

Overall, the simulation results confirm qualitatively the analytical predictions but with important differences. The main deviation is the clear bimodality of the stationary numerical solutions in the presence of bias and for small noise (Fig. \ref{fig_stat_m_diff_w}). While the theoretical prediction for spontaneous magnetization in the thermodynamic limit 
(\ref{eq_magnetization_TD}) presented in Fig. \ref{fig3.1} shows only a single solution corresponding to the global minimum of the free energy, numerical simulations may also converge due to finite-size fluctuations towards metastable states, where at small noise the system dynamics becomes ``trapped'' (see Fig. \ref{fig_avgm_vs_time} and \ref{fig_stat_m_diff_w}a). Corresponding metastable states are consistent with the results obtained from solution of the self consistency equation (Fig. \ref{fig_self_consistency_result}). When we interpret the different branches of solutions in Fig. \ref{fig_self_consistency_result} in the sense of dynamical fixed points of the systems behavior, then red and blue branches correspond to globally and locally stable points, respectively, while the yellow branch should correspond to an unstable point. This interpretation predicts that the numerically obtained distribution of final consensus values must depend on the initial conditions of the system, which is indeed the case: For an initially unbiased system with $m(t=0)=0$, we observe a bimodal distribution with only a minority of runs converging to the metastable state ($+1$ in Fig. \ref{fig_stat_m_diff_w}a). On the other hand, for the same parameters a system initialized in a consensus state $m=+1$, we observe that at low noise all the simulations remain trapped in the metastable state. With increasing noise  the metastable state is predicted to vanish (Fig. \ref{fig_self_consistency_result}). Indeed, at some critical noise value, we observe a jump from a finite, positive average opinion state, to a negative opinion state. For a further increase of $p_{noise}$, we then observe an approach to the undecided state $m=0$ from below (Fig. \ref{fig_stat_m_diff_w}c,d).

\begin{figure}[t!]
    \centering
    \includegraphics[width=1\textwidth]{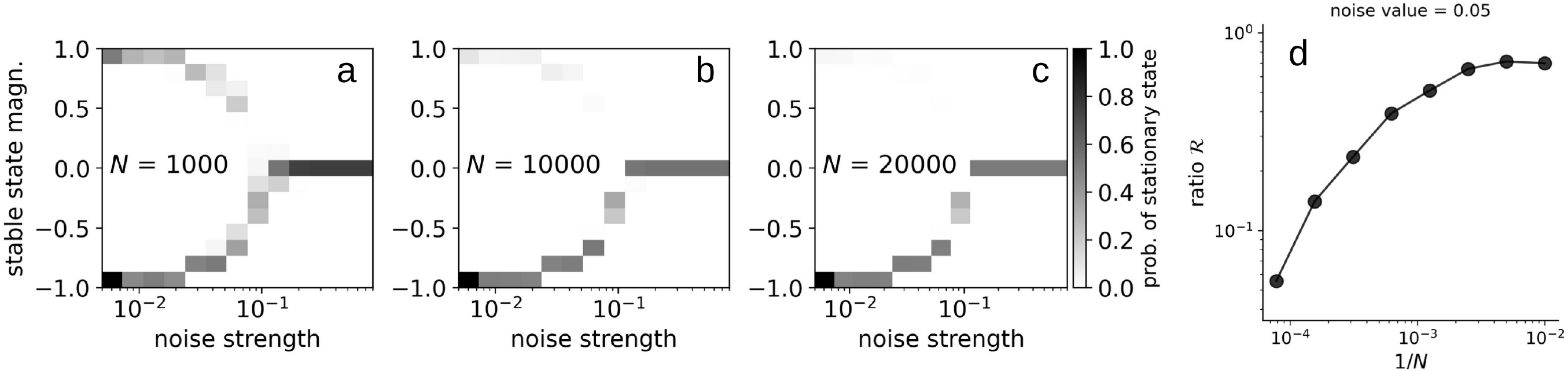}\hspace{3em}
    \caption{
    \textbf{a-c}. Numerical results obtained using the same parameters as the heat-map in Figure~\ref{fig6} for three different system sizes $N \in [1000, 10000, 20000]$.
    \textbf{d}. Numerical results of the ratio $\mathcal{R}$ as a function of the inverse of the system size $1/N$ for eight different system sizes $N \in [100, 200, 400, 800, 1600, 3200, 6400, 12800]$. The rest of the parameters are the same as the ones of the right panel of Figure~\ref{fig6} for only one noise strength value of 0.005.
    }
    \label{fig7}
    \label{fig_sim_increasingN}
\end{figure}

Finally, we test the assumption that the observed deviations reported above are indeed due to finite-size fluctuations preventing the system to converge to the global minimum of free energy. We simulate a system with a bias to the negative opinion for increasing system sizes with the $m=0$ initial condition. As can be seen in Fig. \ref{fig_sim_increasingN}a for small systems ($N=1000$) we observe a relatively high probability for the simulations to converge to the metastable state (positive opinion for $p_{nois}<0.1$), which however decreases with system size, and eventually for $N=20000$ practically vanishes.  
To make it more quantitative, we define the ratio $\mathcal{R} = \mathcal{N}_+/\mathcal{N}_-$, where $\mathcal{N}_+$ is the number of realizations where the steady-state magnetization is positive (metastable state) and $\mathcal{N}_-$ is the number of realizations where the steady-state magnetization is negative (global minimum of the free energy). For a system biased to the negative opinion, as discussed above, a value of $\mathcal{R}=0$ corresponds to all simulations converging to the (global) minimum of the free energy, while a diverging ratio ($\mathcal{R}\to \infty$) would correspond to all simulations being trapped in the metastable state. In Fig. \ref{fig_sim_increasingN}b we show $\mathcal{R}$ as a function of the inverse system size $1/N$, and we observe that the curve approaches the origin for decreasing $1/N$ (increasing system size), which shows that in the thermodynamic limit ($N\to\infty$), we will observe only the results corresponding to the global minimum of the free energy as predicted by theory in Fig \ref{fig_m_vs_T_TD}.

\section{Conclusions and outlook}\label{V}

The typical approach in statistical physics is to formulate a model in terms of a Hamiltonian, derive for it analytical results in the thermodynamic limit, and then to test the analytical predictions with numerical simulations by implementing a dynamic algorithm consistent with the initially formed Hamiltonian. However, when physics inspired spin models are used to study collective decision making and opinion dynamics, they are typically formulated in terms of dynamical models, agent-based models \cite{castellano2009statistical,hartnett2016heterogeneous,winklmayr2020wisdom}. Thus, here we follow partly a ``reverse'' approach: Starting from an agent-based model for collective decision making of agents with heterogeneous preferences, originally introduced by Hartnett \emph{et al.} and previously studied numerically on lattices \cite{hartnett2016heterogeneous}, our aim was to formulate a many-particle Hamiltonian and partition functions and investigate possible analytical solutions. 

We consider two different description of social interactions between individuals with a bias: First, the biased Ising model (bi-model), a superposition of pairwise interactions with the individual bias increasing or decreasing the interaction strength with neighbors holding preferred or disliked opinion, respectively (Sec. \ref{II.2}).  Second, biased agents responding to a local social field (sf-model) generated by its neighbors (Sec. \ref{II.3}). While these approaches are intuitive and straight forward to justify, they are certainly not the only two ways that can be taken. However, both models that we consider are exactly solvable in the case of a complete graph topology of the agent interaction network. With this we are able to single out effects inherent to an all-to-all coupling from those induced by specific network structure as e.g. 2D lattice, considered in the original paper by Hartnett \emph{et al.}. Analytically, we restrict ourselves here to the discussion of stationary states, leaving the questions of dynamics and relaxation towards the steady state for future work. 

The bi-model (Sec. \ref{II.2}) does not exhibit metastable states in the thermodynamic limit ($N\to\infty$). By the steepest descent calculations they vanish in this limit. In contrast to that, in numerical simulations with finite $N$, we can observe a finite probability of individual simulation runs to convergence to a metastable state from a zero-magentization initial condition. However, in agreement with the theoretical prediction, the probability of observing metastable states decreases with increasing system size $N$, and eventually vanishes for sufficiently large $N$ (Fig. \ref{fig_sim_increasingN}). 

In the  sf-model (Sec. \ref{II.3}) the steady state of the magnetization can be obtained from the self-consistency relation. Here, we can also identify solutions corresponding to the metastable states which can be observed in  the numerical simulations at finite $N$. The theory predicts for example the disappearance of the metastable state with increasing temperature corresponding to a saddle-node bifurcation, Fig.\ref{fig_self_consistency_result}, which are linked to the possibility of sudden jumps (discontinuous) in the average magnetization. We were able to directly confirm these predictions in our numerical simulations. In Fig. \ref{fig_stat_m_diff_w}c,d, we show the result for the steady state (average) magnetization for initial conditions strongly favoring relaxation to the metastable state. Here, despite an overall bias towards the negative option at low noise (low temperature) we observe an average positive magnetization, corresponding to a metastable state. However, at a critical noise value we see a discontinuous jump of the magnetization from the positive (metastable) average opinion to a negative one, which corresponds to the globally stable solution. 
This phenomenon could be potentially relevant for opinion dynamics in real-world social systems. For example one can imagine a population of agents that were initially unbiased, or biased towards option $+1$, reaching steady state consensus with $m>0$. As long as the perturbations (noise) are small, even if the preferences of the agents shift slowly towards an overall bias to the negative option $-1$, e.g. by previously unbiased individuals assuming a negative bias, the average opinion will remain ``locked'' to the positive one. However, a change in noise level, or a sufficiently large perturbation, can then exhibit a sudden shift of the average opinion towards the negative consensus opinion, aligned with the underlying negative bias of the agent population. Such tipping-points in social and socio-ecological systems are being widely discussed \cite{castellano2009statistical,centola2018experimental}, and have received attention for example in the context of climate action and sustainability transition \cite{otto2020social,winkelmann2022social}.    

Considering the complete graph allows for exact analytical solutions, but it can not account for effects of particular network structure. The most important difference with respect to the original model on a 2D lattice {\cite{hartnett2016heterogeneous}}, is that on a complete graph, we always observe full consensus due to lack of spatial structure allowing (random) local aggregations of individuals with the same bias to self-reinforce and ``shield'' themselves against a majority opinion, preventing full consensus. On the 2D lattice, the density of unbiased individuals was shown to crucially important to facilitate consensus by breaking up locked in spatial domains of opposite opinion. On a complete graph, the density of unbiased individuals modulates the overall bias in the system and thus the equilibrium magnetization. However, it also controls the structure and stability of metastable solutions as determined via the self-consistency approach of the social field model.

In this study, we have demonstrated how two different Hamiltonians can be formulated based on reasonable assumptions about the nature of social interactions, for a given behavioral algorithm of collective decision making. We have obtained exact steady-state solutions for both Hamiltonians on an``all-to-all" interaction network, using different analytical methods. Our work highlights the power of analytic methods rooted in statistical physics to provide a deep understanding of complex social dynamics. 
While the results obtained on a complete graph can be expected to be similar to the steady state of the system on random networks (Erd\H{o}s-R\'enyi graphs), other network topologies may yield different results. Therefore, our results provide a solid starting point for future investigations of the dynamical behavior of the system, such as convergence to a steady state, and the impact of complex network topologies, including lattices, small-world, or scale-free networks, resembling real-world cases. 

\section*{Acknowledgements}
We acknowledge support by Abel Jonen in piloting/testing individual-based model simulations.
{This work was supported by the BMBF Bridge2ERA program, projekt 01DK20044 (‘Complex networks: self-organization and collective information processing’); Deutsche Forschungsgemeinschaft (DFG, German Research Foundation) under Germany’s Excellence Strategy—EXC 2002/1 ‘Science of Intelligence’, project 390523135 (LG-N and PR); National Academy of Sciences of Ukraine,  project
KPKBK 6541030 (PS, MK, and YuH).}
YuH  acknowledges  useful  discussions  with Yuri Kozitsky (Lublin). MK and YuH acknowledge the hospitality of members of Pawel Romanczuk lab when staying at the Humboldt University Berlin.

\section*{References}
\bibliographystyle{ieeetr}
\bibliography{references.bib}










\end{document}